# On the Pseudo-Schrödinger Equation approximation of the Transfer-Integral operator for 1-dimensional DNA models

Marc JOYEUX

*Laboratoire de Spectrométrie Physique (CNRS UMR 5588),*
*Université Joseph Fourier - Grenoble 1*
*BP 87, 38402 St Martin d'Hères, France*
Marc.Joyeux@ujf-grenoble.fr

*Abstract :* The Transfer-Integral (TI) operator is a powerful method to investigate the statistical physics of 1-dimensional models, like those used to describe DNA denaturation. At the cost of a certain number of approximations, the TI equation can be reduced to a Pseudo-Schrödinger Equation (PSE), according to which the DNA sequence is equivalent to a point particle moving in a potential well. In this paper, I check the validity of the standard PSE approximation for two different 1-dimensional DNA models, and show that it fails to provide correct results for both of them. I then propose a generalized PSE, which works well for one of the two models. Finally, I discuss the particle description of DNA denaturation that is derived from this generalized PSE.

# 1. Introduction

As is well known, DNA consists of two entangled long polymers of simple units. The polymers are called the strands and the monomer units the nucleotides. Each monomer consists of a phosphate group, a sugar group, and a base. The backbones of the strands are made of alternating phosphate and sugar groups connected by ester bonds. A base is attached to each sugar. There are four different types of bases, namely cytosine (C) and thymine (T), which are monocyclic, and guanine (G) and adenine (A), which are bicyclic. The genetic information is encoded in the succession of the various A, T, G and C bases that constitute a specific DNA sequence. The corresponding list of A, T, G and C letters is known as the primary structure (or genome) of the DNA. At physiological temperatures, DNA is essentially observed in the double-stranded structure, which results, as shown by Watson and Crick, from the hydrogen bonds that are formed selectively between A and T and between G and C. The association of the two strands due to base pairing is commonly referred to as the secondary structure of DNA. In addition, DNA sequences also have well-defined higher order conformations, like the B-, A-, and Z- double helix forms, and they eventually further combine with histone proteins to form chromatin. In this paper, we are however only interested in the secondary structure of DNA and not in higher order conformations.

The two strands of DNA separate upon heating [1-4], a phenomenon called "denaturation" or "melting". Homogeneous DNA sequences denaturate abruptly at a precise temperature, while the denaturation of inhomogeneous sequences occurs through a series of steps, which can be monitored by UV absorption spectroscopy [5,6]. At each step, large portions of the inhomogeneous DNA sequence separate over narrow temperature intervals, so that the whole denaturation process looks like a chain of sharp, first order-like phase transitions.

Following the pioneering work of Poland and Scheraga [7,8], the simplest models for denaturation assume, in analogy with the Ising model, that a base pair is either open or closed and that its evolution can be depicted by a two-state variable (see for example Refs. [9-14]). Some of these statistical models are still used nowadays to compute rapidly denaturation curves of long DNA sequences that are in good agreement with experimental results.

While such statistical models were specifically derived to describe DNA melting, dynamical models (that is, models based on a Hamiltonian function of continuous variables) are in principle able to describe the whole dynamics of DNA, from small amplitude oscillations at low temperature to large amplitude motions close to denaturation. The first dynamical model was developed to investigate the properties of solitons in the DNA double strand. In 1980, hydrogen-deuterium exchange experiments indeed evidenced the propagation of base pair openings along the chain in a manner that resembles that of solitons [15]. In the same work, the authors proposed a Hamiltonian for describing DNA, in which the degrees of freedom are the rotation angles of the bases around the strand axes. This first model is quite simple, but more complex models have been proposed since that time and the corresponding solitonic solutions have been investigated (see for example Refs. [16-21]). These models are, however, not aimed at describing DNA denaturation, and cannot be used to investigate it, because denaturation involves essentially the stretching of the base pairs rather than the rotation of the bases around the strands.

To the best of my knowledge, the first dynamical model of DNA that depends on the distance between paired bases and might therefore be expected to describe DNA denaturation correctly was proposed by Prohofsky and co-workers [22,23]. However, Dauxois, Peyrard and Bishop later realized that this original model predicts a denaturation transition that is much too smooth compared to experiments [24,25]. They furthermore showed that the use of an anharmonic stacking potential instead of a harmonic one leads to denaturation curves that are in much better agreement with experiments [26-28]. More recently, other variants of the model of Prohofsky and co-workers were proposed and shown to display the correct sharp behaviour at denaturation (see for example Refs. [29,30]).

All the dynamical models discussed above are mesoscopic 1-dimensional models, in the sense that they describe DNA as a ladder, whose rungs are the paired bases, and one coordinate is sufficient to describe the relative motion of paired nucleotides. Of course, there exist more elaborate models. For example, the model proposed in Ref. [31] is very appealing. This is indeed a mesoscopic model (6 sites, that is 18 coordinates, are used to model one base pair and the associated sugar and phosphate groups), so that standard workstations are sufficient to investigate DNA dynamics at relatively long times. Still, and in contrast with one-dimensional models, the double helix (tertiary structure) is properly taken into account (including the existence of the minor and major grooves) and the molecule is free to deform and diffuse in a three-dimensional buffer. However, the principal drawback of these more realistic models is that the only method that can be used to investigate their dynamics consists in integrating the equations of motion step by step. In contrast, the statistical physics of the 1-dimensional models can be investigated using the powerful technique of the Transfer-Integral (TI) operator method.

The TI operator method is precisely the central point of this paper. As will be described in more detail in Sec. 2, this technique simplifies drastically the evaluation of the partition function - as well as other quantities, like the static form factor - of systems with a large number of degrees of freedom through the replacement of the calculation of multiple integrals involving coupled integrands by that of products of 1-dimensional integrals (see for example Refs. [32-34]). This technique works only for 1-dimensional systems with nearest-neighbour interactions, but the model of Prohofsky and co-workers and other related ones fall precisely in this category. Quite interestingly, application of this method is not limited to infinitely long homogeneous sequences but extends instead to finite [34,35] and inhomogeneous [34,36] ones. The principal advantages of the TI method compared to the step by step integration of the equations of motion are that it is orders of magnitude faster and that it provides correct results closer to the critical temperature (see Fig. 4 of Ref. [37]), because it is not affected by the slow fluctuations that take place in this temperature range [38].

An efficient algorithm to solve the TI equation numerically has been proposed by Schneider and Stoll [33]. The principal step of this algorithm involves the diagonalization of a matrix obtained by replacing the integrals in the TI equation by sums of discrete increments, which yields the eigenvalues of the TI equation, as well as the values of its eigenvectors on the grid. This method is expected to work for all temperatures and coupling strengths (see however the restrictions mentioned in Refs. [34] and [37]). Yet, there exists another method, which consists in replacing, at least approximately, the TI equation by a Pseudo-Schrödinger Equation. This later method has the advantage that it may lead to analytical results. It has consequently been applied to a variety of systems in the low temperature and strong coupling regime where it is expected to be valid (see Refs. [32,39-43] for some early works).

The Pseudo-Schrödinger Equation (PSE) approximation of the TI operator has also been used to interpret qualitatively the behaviour of DNA models at denaturation [24,25,44,45]. Still, for these models very poor agreement is generally found when comparing results obtained from the PSE approximation to exact ones (see for example Fig. 1 of Ref. [25]). The purpose of this paper is to understand why the PSE is so bad an approximation for 1-dimensional DNA models. More precisely, reduction of the TI equation to the PSE involves a series of approximations. I will determine which approximation step is wrong for two different 1-dimensional DNA models, namely the Peyrard-Bishop (PB) model [22-25], and the model that was proposed by our group [29,46], which will hereafter be called the JB model. I will then check whether the PSE approach works correctly when the incriminated approximation is discarded.

The remainder of the paper is organized as follows. The TI operator method and the various approximations leading to the PSE are described in some detail in Sec. 2. I next discuss in Sec. 3 the reasons why the standard PSE approach fails for the PB model, and propose a generalized PSE that works well for this model. Sec. 4 deals with the application of the PSE approach to the JB model. This model is qualitatively different from the PB one, in the sense that it predicts a sharp denaturation instead of a smooth one, in agreement with experimental results. It will be shown that for this later model the PSE approach cannot be adapted and essentially gives meaningless results. Finally, I discuss in Sec. 5 the particle description of DNA denaturation that is derived from the PB model using the generalized PSE.

## 2. The TI Operator Method and the PSE Approximation

In this section, I describe in some detail the TI operator method and the successive approximation steps leading to the PSE.

### 2.1. *The TI operator method*

Let us consider a system that may be described by a set of coordinates $y_k$ ($k=1,2,...,N$) and a potential energy function $E_{\text{pot}}$ of the form

$$E_{\text{pot}} = \sum_{k=1}^{N} V(y_k) + \sum_{k=2}^{N} W(y_{k-1}, y_k) \ , \qquad (2.1)$$

In the case of the 1-dimensional DNA models considered in this paper, $y_k$ represents a measure of the distance between the bases of the *k*th pair, $V(y_k)$ the pairing potential that tends to keep these bases separated by a fixed distance, and $W(y_{k-1}, y_k)$ the stacking potential that models the interactions between neighbouring base pairs. For the sake of simplicity, we consider in this paper only homogeneous sequences, so that functions *V* and *W* do not depend on *k* (they are the same for all base pairs), but it has been shown in Refs. [34,36] how the TI method can be adapted to the case of inhomogeneous sequences. The explicit expressions for $V$ and $W$ are model-dependent and will be provided in Sec. 3 and 4 for the PB and JB models, respectively.

The classical canonical partition function of the system described by the potential energy of Eq. (2.1) is

$$Z = \int \exp[-\beta E_{pot}] \, dy_1 \, dy_2 \ldots dy_N \,, \qquad (2.2)$$

where $\beta = 1/(k_B T)$ is the inverse temperature. The TI method is a technique that allows for the efficient computation of $Z$ (as well as other quantities that can be expressed with similar integrals, like the statistical average of any function $h(y_k)$ and the static form factor). The first step for calculating $Z$ with this method consists in rewriting Eq. (2.2) in the form

$$Z = \int \exp[-\frac{\beta V(y_1)}{2}] K(y_1, y_2) K(y_2, y_3) \ldots K(y_{N-1}, y_N) \exp[-\frac{\beta V(y_N)}{2}] \, dy_1 \, dy_2 \ldots dy_N \,, \qquad (2.3)$$

where the TI kernel $K(y_{k-1}, y_k)$ for base pair $k$-1 interacting with base pair $k$ has the form

$$K(y_{k-1}, y_k) = \exp[-\beta(\frac{V(y_{k-1})}{2} + \frac{V(y_k)}{2} + W(y_{k-1}, y_k))] \,. \qquad (2.4)$$

This kernel is symmetric ($K(y, x) = K(x, y)$) and strictly positive $K(x, y) > 0$, but this is not a Hilbert-Schmidt type one if the coordinates are allowed to vary from -∞ to +∞, because the integral $\iint K^2(x, y) \, dx \, dy$ diverges. For all numerical purposes, one however has to set up a lower and an upper bound for the $y_k$, which amounts to limiting the kernel on a finite subspace. Whatever the bounds for the $y_k$, the norm of the kernel on the subspace is finite, so that the kernel limited on the subspace is of the Hilbert-Schmidt type. It can therefore be expanded in an orthonormal basis [47]

$$K(y_{k-1}, y_k) = \sum_{i=1}^{+\infty} \lambda_i \phi_i(y_{k-1}) \phi_i(y_k) \,, \qquad (2.5)$$

where the $\{\lambda_i\}$ and $\{\phi_i\}$ are the eigenvalues and eigenvectors of the integral operator, which satisfy the TI equation

$$\int K(x, y) \phi_i(y) \, dy = \lambda_i \phi_i(x) \qquad (2.6)$$

in addition to the orthonormality relation

$$\int \phi_i(y) \phi_j(y) \, dy = \delta_{ij} \,. \qquad (2.7)$$

Note that, since the validity of the expansion of Eq. (2.5) depends on the definition of bounds for the $y_k$, one must investigate the case where these bounds become infinite and check that the obtained results do not depend on the precise values of the bounds, provided they are chosen to be sufficiently large [34,37].

By substituting the kernel expansion of Eq. (2.5) in Eq. (2.3) and taking Eq. (2.7) into account, one obtains that $Z$ can be rewritten in the simple form

$$Z = \sum_{i=1}^{+\infty} A_i^2 \lambda_i^{N-1} \;, \tag{2.8}$$

where

$$A_i = \int \phi_i(y) \exp[-\frac{\beta}{2} V(y)] dy \;. \tag{2.9}$$

Since the free energy per base pair, $f$, is defined by

$$f = -\frac{1}{N\beta} \ln[Z] \;, \tag{2.10}$$

one easily sees that, in the limit of infinitely long chains ($N \to +\infty$), $f$ reduces to

$$f = -\frac{1}{\beta} \ln[\lambda_1] \;, \tag{2.11}$$

where $\lambda_1$ is the largest eigenvalue of the integral operator.

From the practical point of view, I used the numerical method of Schneider and Stoll [27,33,46] to compute the $\{\lambda_i\}$ and the values assumed by the $\{\phi_i\}$ on a chosen grid of points. The free energy per base pair, $f$, was then computed according to Eq. (2.10), while closely related quantities, like the entropy per base pair, $s$,

$$s = -\frac{\partial f}{\partial T} \tag{2.12}$$

and the specific heat per base pair, $c_V$,

$$c_V = -T \frac{\partial^2 f}{\partial T^2} \tag{2.13}$$

were estimated from finite differences.

### 2.2. *The PSE approximation of the TI equation*

The PSE may be obtained when trying to solve the TI equation (2.6) without resorting to the discretization on which the numerical method of Schneider and Stoll relies. For this purpose, one first defines

$$\psi_i(y) = \exp[-\frac{\beta}{2}V(y)]\phi_i(y) , \qquad (2.14)$$

which enables to rewrite the TI equation in the explicit form

$$\int \exp[-\beta W(x,y)]\psi_i(y)\,dy = \lambda_i \exp[\beta V(x)]\psi_i(x) . \qquad (2.15)$$

At this point, one may take advantage of the fact that, for many models, the stacking potential $W(x,y)$ is actually an even function of $z = y - x$,

$$W(x,y) = W(y-x) = W(z) , \qquad (2.16)$$

which increases rapidly as $z$ departs from 0. Then, by replacing in Eq. (2.15) $\psi_i(y)$ by its Taylor series expansion around $y = x$, one can rewrite the TI equation in the form

$$\sum_{p \geq 0} a_p \psi_i^{(2p)}(x) = \lambda_i \exp[\beta V(x)]\psi_i(x) , \qquad (2.17)$$

where

$$a_p = \frac{1}{(2p)!}\int_{-\infty}^{+\infty} z^{2p}\exp[-\beta W(z)]\,dz . \qquad (2.18)$$

Note that this involves a first approximation, because it is not granted that the Taylor series expansion of $\psi_i(y)$ is convergent everywhere. At some point, it may therefore be necessary to check that the resolution of Eqs. (2.15) and (2.17) leads to identical values of the $\{\lambda_i\}$ and the $\{\phi_i\}$.

The approximations to come are, however, much more drastic than this first one. Indeed, the second approximation consists in considering that

$$\frac{\lambda_i}{a_0}\exp[\beta V(x)] = \exp[\beta V(x) + \ln[\frac{\lambda_i}{a_0}]] \approx 1 + \beta V(x) + \ln[\frac{\lambda_i}{a_0}] \qquad (2.19)$$

and in replacing Eq. (2.19) in Eq. (2.17).

At last, the third (and last) approximation consists in retaining, in the left-hand side of Eq. (2.17), only the first two terms of the expansion, that is, the terms with $p=0$ and $p=1$. As a result, one finally gets the PSE

$$-\frac{1}{2m}\psi_i^{(2)}(x) + V(x)\psi_i(x) = \tilde{\varepsilon}_i \psi_i(x) , \qquad (2.20)$$

where

$$m = \frac{\beta a_0}{2a_1}$$

$$\tilde{\varepsilon}_i = -\frac{1}{\beta}\ln(\frac{\lambda_i}{a_0}) \ . \tag{2.21}$$

Needless to say, the second and third approximations must also be considered with some caution.

## 3. Application to the Peyrard-Bishop (PB) model

In this section, I first show that the standard PSE approximation in Eq. (2.20) fails to provide correct results even for the very simple PB model [22-25]. Then, I point out that rather accurate results are instead obtained when using Eq. (2.17) instead of Eq. (2.20).

### 3.1. *The PB model*

In the PB model [22-25], the pairing potential $V$ is represented by a Morse potential

$$V(\delta r_k) = D(1-\exp[-a\,\delta r_k])^2 = D + D(\exp[-2a\,\delta r_k] - 2\exp[-a\,\delta r_k]) \ , \tag{3.1}$$

where $\delta r_k$ is the deviation from its equilibrium value of the distance between the bases of the $k$th pair, while the stacking interaction between successive base pairs, $W$, is assumed to be harmonic

$$W(\delta r_{k-1}, \delta r_k) = \frac{K}{2}(\delta r_k - \delta r_{k-1})^2 \ . \tag{3.2}$$

In this work, I used the numerical values of the parameters reported in Ref. [48], that is, $D$=0.063 eV, $a$=4.2 Å$^{-1}$ and $K$=0.025 eV Å$^{-2}$.

However, the functions $\psi_i$ must be dimensionless if they are to be considered as wave functions. This implies, in turn, that the coordinates $y_k$ in Eq. (2.1) must also be dimensionless. One therefore defines

$$y_k = a\,\delta r_k \tag{3.3}$$

and rewrites $V$ and $W$ in the form

$$V(y_k) = D(1-\exp[-y_k])^2 = D + D(\exp[-2y_k] - 2\exp[-y_k])$$

$$W(y_{k-1}, y_k) = \frac{K}{2a^2}(y_k - y_{k-1})^2 \ . \tag{3.4}$$

### 3.2. *Application of the standard PSE approximation to the PB model*

The coefficients $a_p$ defined in Eq. (2.18) can be evaluated analytically for the harmonic stacking interaction of Eq. (3.4). One gets

$$a_p = \frac{(2p-1)!!}{(2p)!}\sqrt{\frac{\pi}{u}}(2u)^{-p} ,  \quad (3.5)$$

where

$$u = \frac{\beta K}{2a^2} . \quad (3.6)$$

Therefore, one has, from Eq. (2.21),

$$m = \frac{\beta^2 K}{a^2}$$

$$\tilde{\varepsilon}_i = -\frac{1}{\beta}\ln[\lambda_i] + \frac{1}{2\beta}\ln[\frac{2\pi a^2}{\beta K}] . \quad (3.7)$$

Analytical solution of the Schrödinger equation with the potential function $V$ in Eq. (3.4) has been known since the early work of Morse [49,50]. What is most relevant for 1-dimensional DNA models, is that, at sufficiently low temperatures, the Schrödinger equation has a bound ground-state solution with energy

$$\tilde{\varepsilon}_1 = \sqrt{\frac{D}{2m}} - \frac{1}{8m} . \quad (3.8)$$

However, at a critical temperature $T_c$, such that

$$T_c = \frac{\sqrt{8KD}}{ak_B} , \quad (3.9)$$

$m$ becomes equal to $m = 1/(8D)$, so that $\tilde{\varepsilon}_1 = D$. This implies that the wave function of the ground state is no longer confined into the well of the Morse potential but can instead extend to infinity. Since the coordinate $x$ in Eq. (2.20) is proportional to the distance that separates the two strands, this implies that at $T_c$ the DNA sequence denaturates and switches from the double-stranded configuration to the single-stranded one.

When combining Eqs. (2.11), (3.7) and (3.8), one finally gets that the free energy per base pair is equal to

$$f = \sqrt{\frac{D}{2m}} - \frac{1}{8m} - \frac{1}{2\beta}\ln[\frac{2\pi a^2}{\beta K}] \quad (3.10)$$

for the double-stranded configuration.

### 3.3. *Comparison of the standard PSE approximation with exact results*

At this point, it is quite instructive to check the accuracy of results obtained with the standard PSE approximation. For example, Fig. 1 shows the temperature evolution of the free energy per base pair, *f*, according to the PB model. The solid line shows "exact" results estimated directly from the TI equation, while the dashed line shows results obtained from Eq. (3.10). Exact results were computed with the numerical algorithm of Schneider and Stoll [27,33,46] and a grid with irregular spacing, like that used in Refs. [46,51]. This grid consists of 4201 values ranging from about -40 to about 5000. The spacing between two points of the grid increases exponentially from 0.2 at the origin to 4.0 for values close to 5000.

Fig. 1 indicates unambiguously that the PSE results are completely wrong, even at quite low temperatures. The denaturation temperature itself is also estimated with a very large error. Eq. (3.9) indeed leads to $T_c = 310$ K, while the exact critical temperature, estimated as the temperature for which the correlation length $\xi$ computed as in Refs. [29,37,46] is maximum, is close to 556 K.

### 3.4. *What is wrong in the PSE approximation ?*

The question therefore is: what is wrong with the PSE approximation ? A rather natural step for answering this question consists in checking whether Eq. (2.17) leads to correct results or whether the results obtained there with are also wrong. Indeed, if such results are correct, then the replacement in Eq. (2.15) of $\psi_i(y)$ by its Taylor series expansion around $y = x$ is valid and the failure of the PSE approximation is attributable to one of the two subsequent approximations. In the opposite case, the very first step of the reduction of the TI equation to the PSE is already not valid.

We therefore need a reliable method to solve the eigenvalue problem of Eq. (2.17). Since close to the melting temperature $\psi_i(x)$ is large for large values of *x*, one is more or less obliged to use the canonical basis of the harmonic oscillator, that is, the set of functions

$$e_n(q) = \frac{1}{\sqrt{2^n n! \sqrt{\pi}}} \exp[-\frac{q^2}{2}] H_n(q) , \qquad (3.11)$$

where the $H_n(q)$ are the Hermite polynomials. One therefore introduces the dimensionless parameter $\gamma$, such that

$$x = \gamma q, \tag{3.12}$$

as well as the functions $\chi_i(q)$, such that

$$\chi_i(q) = \psi_i(\gamma q) = \psi_i(x), \tag{3.13}$$

and rewrites Eq. (2.17) in the form

$$H\chi_i(q) = \lambda_i \chi_i(q) \tag{3.14}$$

where

$$H = \sum_{p \geq 0} b_p \exp[-\beta V(\gamma q)] \left(\frac{\partial^2}{\partial q^2}\right)^p, \tag{3.15}$$

and

$$b_p = \frac{a_p}{\gamma^{2p}}. \tag{3.16}$$

The matrix representation of the differential operator $\partial^2/\partial q^2$ in the basis of the $e_n(q)$ functions is built using the relation

$$\frac{\partial^2 e_n(q)}{\partial q^2} = \frac{1}{2}\sqrt{(n+1)(n+2)}\, e_{n+2}(q) - (n+\frac{1}{2})e_n(q) + \frac{1}{2}\sqrt{n(n-1)}\, e_{n-2}(q). \tag{3.17}$$

The matrix representation of the operator $H$ is then built from Eq. (3.15), by computing the corresponding linear combination of products of single operator representations, and diagonalized. Several comments are in order at that point.

(a) because of the highly oscillatory nature of the $e_n(q)$ functions, it is difficult to built accurate matrix representations of the operators $\exp[-\beta V(\gamma q)]$ and $H$ with size much larger than about 60 by 60.

(b) the choice of the dimensionless parameter $\gamma$ necessarily results from the balance between several conflicting constraints. Indeed, given the maximum practical size $n \leq n_{max} \approx 60$ of the basis of $e_n(q)$ functions, the larger the value of $\gamma$, the broader the interval of values of $x$ where the wave function $\psi_i(x)$ may assume non-vanishing values, but, on the other hand, the rougher and the more oscillatory the estimated wave functions.

(c) more dramatic from the numerical point of view is the fact that too small values of $\gamma$ lead to $b_p$ series that decrease too slowly with $p$. As a consequence, the off-diagonal elements of the matrix representation of $H$ become too large, and it is no longer possible to get reliable estimates of its eigenvalues, that is, estimates that do not vary significantly when the size $n_{max}$ of the basis is increased.

(d) the operator $H$ is not hermitian, so that not all of its eigenvalues are real.

(e) when the parameter $\gamma$ is chosen large enough, the eigenvalue with largest real part is real. This is an estimate of the largest eigenvalue $\lambda_1$ of the integral operator, which appears for example in the expression of the free energy $f$ (see Eq. (2.11)).

Results obtained by taking these five points into consideration are shown in Figs. 2 to 4. The top plot in each figure shows the temperature evolution of $f$ and the bottom plot that of $c_V$. The thick solid lines show exact results obtained with the TI method, while the other lines show results obtained from Eqs. (3.14) and (3.15) for different values of the parameter $\gamma$, the basis size $n_{max}$, and the order $p_{max}$ at which the series in Eq. (3.15) is truncated.

Fig. 2 shows results obtained with $\gamma = 30$, $p_{max} = 4$, and $n_{max}$ increasing from 30 to 60. As expected, it is seen that results become better with increasing basis size. All subsequent calculations were therefore performed with the maximum practical basis size $n_{max} = 60$.

Fig. 3 shows results obtained with $n_{max} = 60$, $p_{max} = 4$, and $\gamma$ increasing from 20 to 50. This figure illustrates clearly points (b) and (c) discussed above. Indeed, it is seen that the convergence of the plots is uncertain for small values of $\gamma$, while the quality of the results again degrades for larger values of $\gamma$. For the particular model studied here, the optimum value of $\gamma$ is close to $\gamma = 30$.

At last, Fig. 4 shows results obtained with $n_{max} = 60$, $\gamma = 30$, and $p_{max}$ increasing from 1 to 20 (the curves with $p_{max} = 10$ and $p_{max} = 20$ superpose). As expected, again, the best results are obtained for the largest values of $p_{max}$. Except for the fact that they vary more smoothly close to the critical temperature, the curves for $p_{max} = 10$ and $p_{max} = 20$ are rather close to the exact ones. Still, it should be noted that results obtained with $p_{max} = 1$ are already not so bad. Comparison of Fig. 1 with the top plot of Fig. 4 indeed indicates that they are much closer to exact ones than those obtained with the standard PSE approximation of Eq. (3.10). I will come back to this point in Sec. 5.

Conclusion therefore is that, in the case of the PB model, the failure of the standard PSE approximation is essentially due to the approximation of Eq. (2.19) and, to a substantially lesser extent, to the truncation of the series in Eq. (2.17) or Eq. (3.15). In contrast, quite accurate results can still be obtained with the generalized PSE formulation of Eqs. (3.14) and (3.15), especially when accepting to deal with derivative operators with order larger than 2.

## 4. Application to the Joyeux-Buyukdagli (JB) model

In this section, I will check whether the conclusions drawn in the preceding section for the PB model [22-25] also hold for the 1-dimensional DNA model that was recently proposed by our group [29,46].

### 4.1. *The JB model*

Like for the PB model, the pairing potential $V$ of the JB model is represented by the Morse potential of Eqs. (3.1) and (3.4). In contrast, the stacking interaction between successive base pairs, $W$, is assumed to be of the form

$$W(\delta r_{k-1}, \delta r_k) = \frac{\Delta H}{2}(1 - \exp[-b(\delta r_k - \delta r_{k-1})^2]) + K_b(\delta r_k - \delta r_{k-1})^2 \,, \tag{4.1}$$

that is, in terms of the $y_k$,

$$W(y_{k-1}, y_k) = \frac{\Delta H}{2}(1 - \exp[-\frac{b}{a^2}(y_k - y_{k-1})^2]) + \frac{K_b}{a^2}(y_k - y_{k-1})^2 \,. \tag{4.2}$$

The first term in the right-hand side of these equations describes the finite stacking interaction and the second one the stiffness of the phosphate/sugar backbone. It was shown in Refs [29,46,52] that the introduction of finite stacking enthalpies $\Delta H$ in the expression of $W$ is in itself sufficient to ensure a sharp, first-order looking melting phase transition, which is in better agreement with experimental data than the smooth transition of the PB model.

At this point, it is worth noting again that Dauxois, Peyrard and Bishop proposed an anharmonic stacking potential, which differs from Eqs. (4.1) and (4.2) but also leads to sharp melting curves [25-27]. However, their expression for $W$ depends not only on $y_k - y_{k-1}$, but also on $y_k + y_{k-1}$, so that it is no longer possible to rewrite $W$ as an even function of $z = y - x$ as in Eq. (2.16). In order to get a differential expression like Eq. (2.17), one would consequently need to replace in Eq. (2.15) both $\psi_i(y)$ and $W(x, y)$ by their Taylor series expansions around $y = x$, so that the whole procedure would become more questionable and the results even more uncertain. In this section, we thus concentrate on the JB model.

The results presented below were obtained with the numerical values of the parameters reported in Ref. [37], that is, $D=0.048$ eV, $a=6.0$ Å$^{-1}$, $\Delta H =0.818$ eV, $b=0.80$ Å$^{-2}$, and $K_b = 4.0 \times 10^{-4}$ eV Å$^{-2}$.

## 4.2. *Failure of the PSE approximation*

Results obtained for the JB model are shown in Fig. 5. As in Figs. 2-4, the top plot shows the temperature evolution of $f$ and the bottom one that of $c_V$. In both plots, the thick solid line shows exact results obtained with the TI method and the numerical algorithm of Schneider and Stoll [27,33,46]. The fact that the JB model leads to a much sharper DNA denaturation transition than the PB one is reflected in the sudden increase of the specific heat close to the critical temperature $T_c \approx 360$ K.

The thinner lines in Fig. 5 show results obtained from Eqs. (3.14) and (3.15). In contrast with the PB model, the coefficients $a_p$ defined in Eq. (2.18), and consequently also the coefficients $b_p$ defined in Eq. (3.16), cannot be evaluated analytically for the anharmonic stacking interaction of Eq. (4.2) and must consequently be estimated numerically. The results shown in Fig. 5 were obtained with a basis size $n_{max} = 50$, an expansion order $p_{max} = 1$, and values of $\gamma$ increasing regularly from 20 to 100. It is seen that these results are essentially meaningless for double-stranded DNA, that is, below the critical temperature $T_c$. Moreover, such results become still worse when the expansion of Eq. (3.15) is truncated at larger values of $p_{max}$ and/or smaller values of $\gamma$ are used.

As already emphasized, the only approximation that is made when going from the TI equation (2.6) to the generalized PSE formulation of Eqs. (3.14) and (3.15) is the replacement in Eq. (2.15) of $\psi_i(y)$ by its Taylor series expansion around $y = x$. The fact that the results obtained from Eqs. (3.14) and (3.15) are so clearly wrong (at any truncation order, and particularly at the order $p_{max} = 1$ that leads to a standard second-order PSE) indicates that this replacement is not valid for the JB model. Since this is the very first one of the series of approximations leading from the TI equation to the PSE, it must be concluded that the PSE approximation can definitely not be used to investigate the dynamics of the JB model.

From the physical point of view, this failure is due to the complex shape of the anharmonic stacking potential $W$ of the JB model, which leads to a much sharper increase of the series of $a_p$ coefficients of Eq. (3.5) than for the PB model. For example, at 300 K, the first terms of this series are $(1.\times 10^1, 1.\times 10^2, 4.\times 10^2, 1.\times 10^3, 3.\times 10^3, 6.\times 10^3, ...)$ for the PB model, and $(8.\times 10^0, 1.\times 10^3, 3.\times 10^5, 5.\times 10^7, 8.\times 10^9, 9.\times 10^{11}, ...)$ for the JB one.

## 5. Conclusion

It has just been shown that the PSE approximation cannot be used to investigate the dynamics of the JB model for DNA. In contrast, it was shown in Sec. 3, that a generalized PSE formulation does provide reasonable results for the smoother PB model. Still, I would like to emphasize in this conclusive section that the description of DNA melting obtained from this generalized PSE formulation is quite different from the one that may be inferred from the standard PSE.

Indeed, according to the standard PSE in Eq. (2.20), DNA sequences may be considered as single particles with a mass $m$ given in Eq. (2.21) moving in a potential, which is just the pairing potential $V$. In the case of the PB model, the mass $m$ is explicitly related to the parameters of the model through Eq. (3.7), while the pairing potential $V$ is the Morse function of Eqs. (3.1) or (3.4). The mass $m$ decreases with increasing temperature like the inverse of $T$. At low temperatures, the particle is very heavy and its ground state is therefore confined deep inside the bottom of the Morse well. As temperature increases, the particle however becomes lighter and lighter, so that its total energy increases and comes closer and closer to the dissociation threshold $D$ of the Morse potential. At the critical temperature $T_c$, the energy of the ground state of the particle is equal to $D$, so that it is no longer confined close to the origin but can instead go to infinity, which amounts to say that the DNA sequence denaturates.

Examination of Fig. 1 indicates that this rather frequently proposed description of DNA melting (as well as other closely related systems [53]) fails to capture the complexity of behaviour of DNA in the terms considered here. The description obtained from the generalized PSE formulation is, indeed, substantially more complex. When truncated at second order ($p_{max} = 1$), Eq. (2.17) may be rewritten in the form

$$-\frac{1}{2m_{eff}(x)}\psi_i^{(2)}(x) + V_{eff}(x)\psi_i(x) = \mu_i \psi_i(x) ,  \quad (5.1)$$

where

$$m_{eff}(x) = m\exp[\beta V(x)]$$
$$V_{eff}(x) = \frac{1}{\beta}(1-\exp[-\beta V(x)]) \quad (5.2)$$
$$\mu_i = \frac{1}{\beta}(1-\frac{\lambda_i}{a_0})$$

and $m$ is given in Eq. (2.21). Comparison of Eqs. (2.20) and (2.21) with Eqs. (5.1) and (5.2) indicates that $m$ and $V(x)$ are the lowest order approximations of, respectively, $m_{eff}(x)$ and $V_{eff}(x)$ for small values of $\beta V(x)$ (that is, when the pairing energy is small compared to the thermal one), while $\mu_i$ is a low order expansion of $\tilde{\varepsilon}_i$ when $\lambda_i$ is close to $a_0$. None of these two conditions ($\beta V(x) \approx 0$ and $\lambda_1 \approx a_0$) is satisfied when studying DNA denaturation, so that it is not really surprising that the standard PSE approximation totally fails.

Plots of $V_{eff}(x)$ and $m_{eff}(x)$ at $T=100$ K and $T=450$ K are shown in Figs. 6 and 7, respectively. It is seen that the effective potential $V_{eff}(x)$ deepens and widens considerably as temperature increases but that the energy of the DNA "particle" remains close to the threshold for all temperatures. Most interestingly, it moreover appears that denaturation is consequently due to a strong modification of the position dependence of the effective mass $m_{eff}(x)$ with increasing temperatures. It is indeed seen in Fig. 6 that, for small temperatures, the effective mass $m_{eff}(x)$ is orders of magnitude larger for large values of $x$ than for $x=0$, where $m_{eff}(x) = m$. This, of course, tends to localize the particle close to $x=0$, that is, to keep the DNA in the double-stranded configuration. In contrast, for temperatures close to or above the critical one, the effective mass $m_{eff}(x)$ is only a few times larger than $m$ at large values of $x$, so that mass effects no longer prevent DNA from melting.

# REFERENCES

[1] R. Thomas, Structure secondaire et dénaturation des acides nucléiques, *Bull. Soc. Chim. Biol. (Paris)* **35** (1953) 609-614.

[2] R. Thomas, Recherches sur la dénaturation ,des acides desoxyribonucléiques, *Biochim. Biophys. Acta* **14** (1954) 231-240.

[3] J. Marmur, R. Rownd and C.L. Schildkraut, Denaturation and renaturation of deoxyribonucleic acid, *Prog. Nucleic Acid Res. Mol. Biol.* **1** (1963) 231-300.

[4] D. Poland and H.A. Scheraga, *Theory of helix-coil transitions in biopolymers* (Academic Press, New York, 1970).

[5] O. Gotoh, Prediction of melting profiles and local helix stability for sequenced DNA, *Adv. Biophys*. **16** (1983) iii.

[6] R.M. Wartell and A.S. Benight, Thermal denaturation of DNA molecules : a comparison of theory with experiment, *Phys. Rep.* **126** (1985) 67-107.

[7] D. Poland and H.A. Scheraga, Phase transitions in one dimension and the helix-coil transition in polyamino acids, *J. Chem. Phys.* **45** (1966) 1456-1463.

[8] D. Poland and H.A. Scheraga, Occurence of a phase transitions in nucleic acid models, *J. Chem. Phys.* **45** (1966) 1463-1469.

[9] D. Poland, Recursion relation generation of probability profiles for specific-sequence macromolecules with long-range correlations, *Biopolymers* **13** (1974) 1859-1871.

[10] M. Fixman and J.J. Freire, Theory of DNA melting curves, *Biopolymers* **16** (1977) 2693-2704.

[11] J. SantaLucia, A unified view of polymer, dumbbell, and oligonucleotide DNA nearest-neighbor thermodynamics, *Proc. Natl. Acad. Sci. USA* **95** (1998) 1460-1465.

[12] R.D. Blake and S.G. Delcourt, Thermal stability of DNA, *Nucleic Acids Res.* **26** (1998) 3323-3332.

[13] R.D. Blake, J.W. Bizzaro, J.D. Blake, G.R. Day, S.G. Delcourt, J. Knowles, K.A. Marx and J. SantaLucia, Statistical mechanical simulation of polymeric DNA melting with MELTSIM, *Bioinformatics* **15** (1999) 370-375.

[14] R. Blossey and E. Carlon, Reparametrizing the loop entropy weights : effect on DNA melting curves, *Phys. Rev. E* **68** (2003) 061911.

[15] S.W. Englander, N.R. Kallenbach, A.J. Heeger, J.A. Krumhansl and S. Litwin, Nature of the open state in long polynucleotide double helices: possibility of soliton excitations, *Proc. Natl. Acad. Sci. USA* **77** (1980) 7222-7226.

[16] S. Yomosa, Solitary excitations in deoxyribonucleic-acid (DNA) double helices, *Phys. Rev. A* **30** (1984) 474-480.

[17] L.V. Yakushevich, Nonlinear DNA dynamics - a new model, *Phys. Lett. A* **136** (1989) 413-417.

[18] G. Gaeta, Solitons in planar and helicoidal Yakushevich model of DNA dynamics, *Phys. Lett. A* **168** (1992) 383-390.

[19] G. Gaeta, A realistic version of the Y-model for DNA dynamics and selection of soliton speed, *Phys. Lett. A* **190** (1994) 301-308.

[20] G. Gaeta, Solitons in the Yakushevich model of DNA beyond the contact approximation, *Phys. Rev. E* **74** (2006) 021921.

[21] M. Cadoni, R. De Leo and G. Gaeta, Composite model for DNA torsion dynamics, *Phys. Rev. E* **75** (2007) 021919.

[22] Y. Gao and E.W. Prohofsky, A modified self-consistent phonon theory of hydrogen bond melting, J. Chem. Phys. 80 (1984) 2242-2243.

[23] M. Techera, L.L. Daemen and E.W. Prohofsky, Nonlinear model of the DNA molecule, *Phys. Rev. A* **40** (1989) 6636-6642.

[24] M. Peyrard and A.R. Bishop, Statistical mechanics of a nonlinear model for DNA denaturation, *Phys. Rev. Lett.* **62** (1989) 2755-2758.

[25] T. Dauxois, M. Peyrard and A.R. Bishop, Dynamics and thermodynamics of a nonlinear model for DNA denaturation, *Phys. Rev. E* **47** (1993) 684-695.

[26] T. Dauxois, M. Peyrard and A.R. Bishop, Entropy-driven DNA denaturation, *Phys. Rev. E* **47** (1993) R44-R47.

[27] T. Dauxois and M. Peyrard, Entropy-driven transition in a one-dimensional system, *Phys. Rev. E* **51** (1995) 4027-4040.

[28] N. Theodorakopoulos, T. Dauxois and M. Peyrard, Order of the phase transition in models of DNA thermal denaturation, *Phys. Rev. Lett.* **85** (2000) 6-9.

[29] M. Joyeux and S. Buyukdagli, Dynamical model based on finite stacking enthalpies for homogeneous and inhomogeneous DNA thermal denaturation, *Phys. Rev. E* **72** (2005) 051902.

[30] G. Weber, Sharp DNA denaturation due to solvent interaction, *Europhys. Lett.* **73** (2006) 806-811.

[31] T.A. Knotts, N. Rathore, D.C. Schwartz and J.J. de Pablo, A coarse grain model for DNA, *J. Chem. Phys.* **126** (2007) 084901.

[32] D.J. Scalapino, M. Sears and R.A. Ferrell, Statistical mechanics of one-dimensional Ginzburg-Landau fields, *Phys. Rev. B* **6** (1972) 3409-3416.

[33] T. Schneider and E. Stoll, Classical statistical mechanics of the sine-Gordon and $\phi^4$ chains. Static properties, *Phys. Rev. B* **22** (1980) 5317-5338.

[34] Y.-L. Zhang, W.-M. Zheng, J.-X. Liu and Y. Z. Chen, Theory of DNA melting based on the Peyrard-Bishop model, *Phys. Rev. E* **56** (1997) 7100-7115.

[35] S. Buyukdagli and M. Joyeux, Theoretical investigation of finite size effects at DNA melting, *Phys. Rev. E* **76** (2007) 021917.

[36] S. Buyukdagli and M. Joyeux, Statistical physics of the melting of inhomogeneous DNA, *Phys. Rev. E* **77** (2008) 031903.

[37] M. Joyeux and A.-M. Florescu, Dynamical versus statistical mesoscopic models for DNA denaturation, *J. Phys.: Condens. Matter* **21** (2009) 034101.

[38] M. Joyeux, S. Buyukdagli and M. Sanrey, 1/$f$ fluctuations of DNA temperature at thermal denaturation, *Phys. Rev. E* **75** (2007) 061914.

[39] J.A. Krumhansl and J.R. Schrieffer, Dynamics and statistical mechanics of a one-dimensional model Hamiltonian for structural phase transitions, *Phys. Rev. B* **11** (1975) 3535-3545.

[40] G.F. Mazenko and P.S. Sahni, Statistical-mechanical treatment of kinks in a one-dimensional model for displacive phase transitions, *Phys. Rev. B* **18** (1978) 6139-6159.

[41] J.F. Currie, J.A. Krumhansl, A.R. Bishop and S.E. Trullinger, Statistical mechanics of one-dimensional solitary-wave-bearing scalar fields: Exact results and ideal-gas phenomenology, *Phys. Rev. B* **22** (1980) 477-496.

[42] N. Gupta and B. Sutherland, Investigation of a class of one-dimensional nonlinear fields, *Phys. Rev. A* **14** (1976) 1790-1801.

[43] J.F. Currie, M.B. Fogel and F.L. Palmer, Thermodynamics of the sine-Gordon field, *Phys. Rev. A* **16** (1977) 796-798.

[44] T. Dauxois, N. Theodorakopoulos and M. Peyrard, Thermodynamic instabilities in one dimension: Correlations, scaling and solitons, *J. Stat. Phys.* **107** (2002) 869-891.

[45] M. Peyrard, Nonlinear dynamics and statistical physics of DNA, *Nonlinearity* **17** (2004) R1-R40.

[46] S. Buyukdagli and M. Joyeux, Scaling laws at the phase transition of systems with divergent order parameter and/or internal length: The example of DNA denaturation, *Phys. Rev. E* **73** (2006) 051910.

[47] A.C. Pipkin, *A course on integral equations* (Springer, Berlin, 1991).

[48] N. Singh and Y. Singh, Statistical theory of force-induced unzipping of DNA, *Eur. Phys. J. E* **17** (2005) 7-19.

[49] P.M. Morse, Diatomic Molecules According to the Wave Mechanics. II. Vibrational Levels, *Phys. Rev.* **34** (1929) 57-64.

[50] G. Chen, The exact solutions of the Schrödinger equation with the Morse potential via Laplace transforms, *Phys. Lett. A* **326** (2004) 55-57.

[51] S. Buyukdagli and M. Joyeux, Mapping between the order of thermal denaturation and the shape of the critical line of mechanical unzipping in one-dimensional DNA models, *Chem. Phys. Lett.* **484** (2010) 315-320.

[52] S. Buyukdagli, M. Sanrey and M. Joyeux, Towards more realistic dynamical models for DNA secondary structure, *Chem. Phys. Lett.* **419** (2006) 434-438.

[53] S. Ares and A. Sanchez, Equilibrium roughening transition in a one-dimensional modified sine-Gordon model, *Phys. Rev. E* **70** (2004) 061607.

**FIGURE 1**

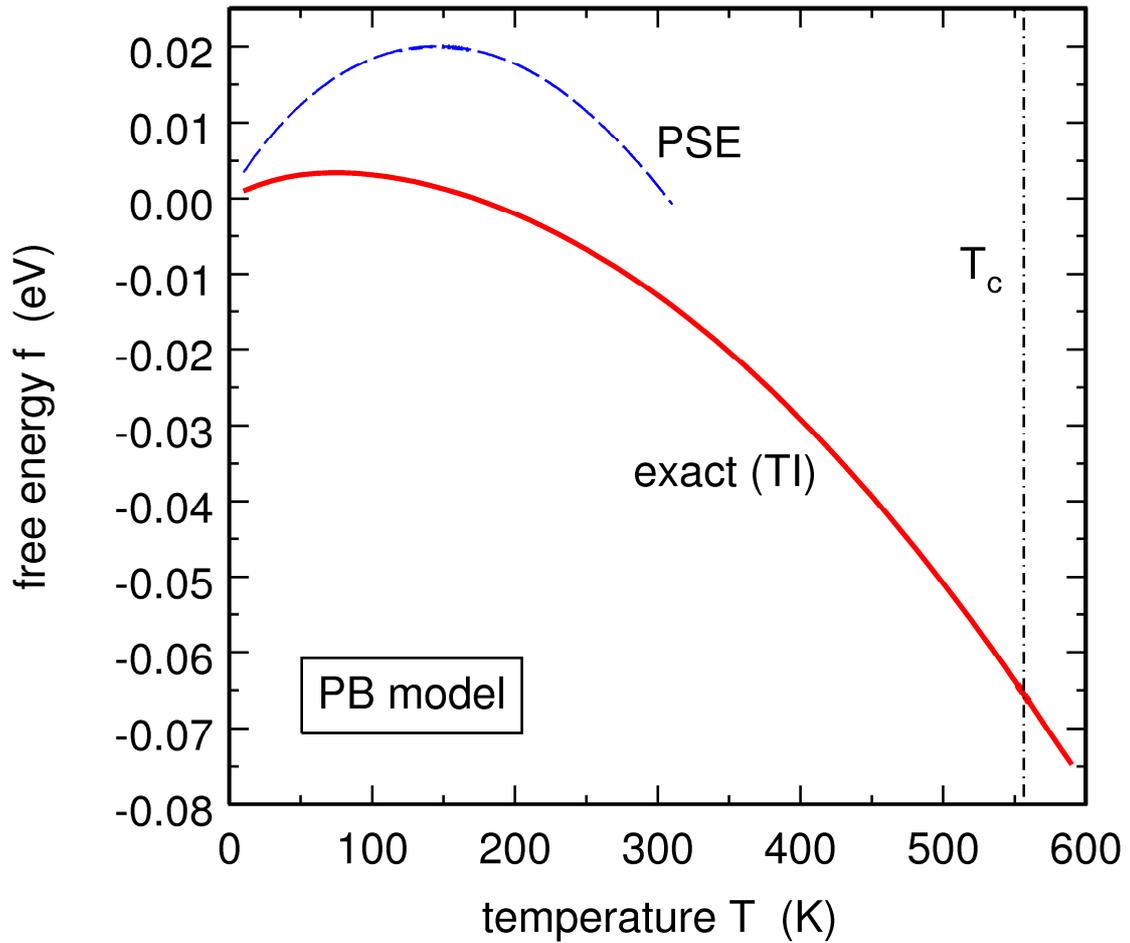

**Fig. 1** : Temperature evolution of the free energy per base pair $f$, expressed in eV, for the PB model. The solid line shows the exact result obtained from the TI equation, while the dashed line shows the result obtained from Eq. (3.10). The vertical dot-dashed line indicates the correct critical temperature $T_c$ determined from the TI equation.

FIGURE 2

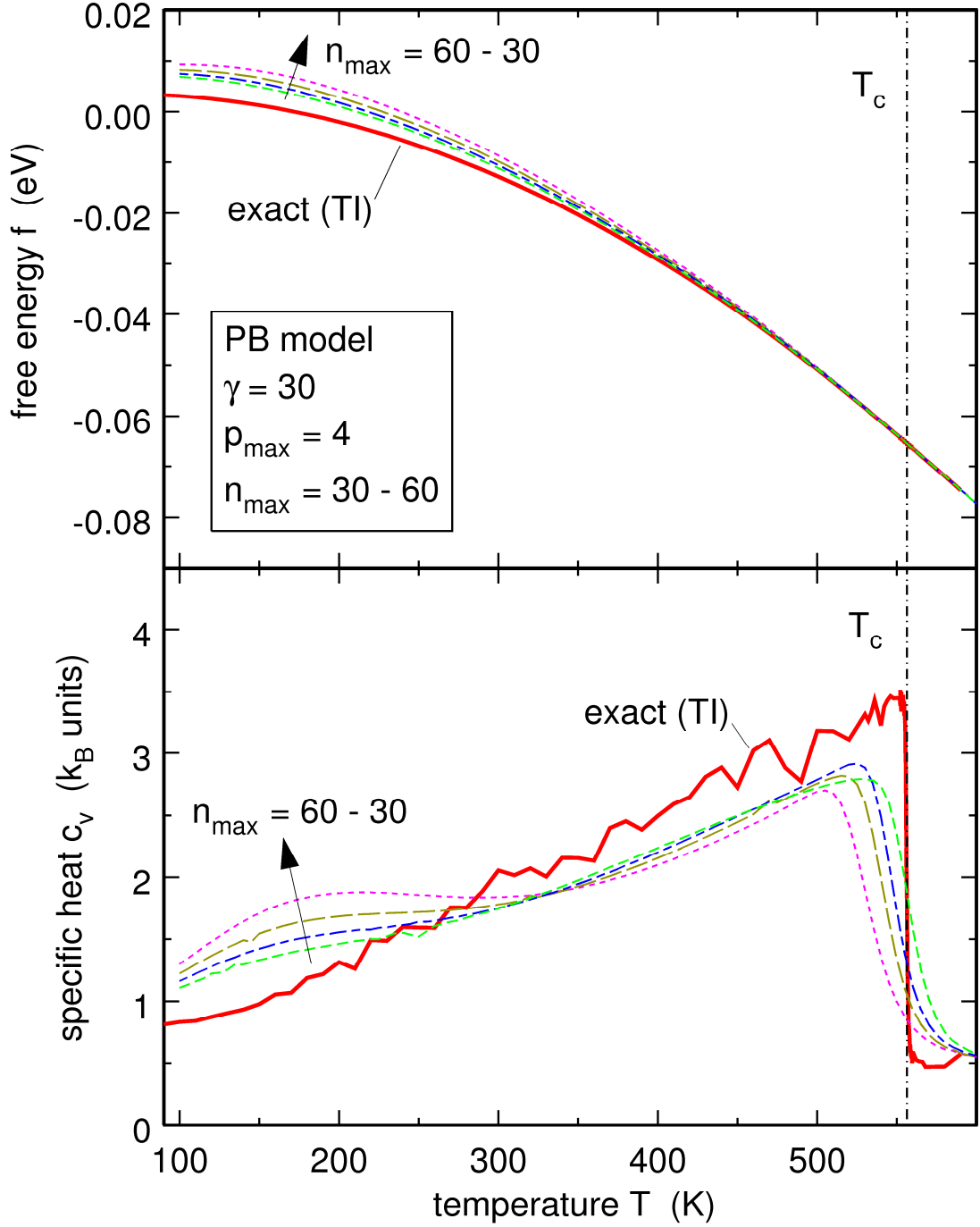

**Fig. 2** : Temperature evolution of $f$, expressed in eV (top plot), and $c_V$, expressed in units of $k_B$ (bottom plot), for the PB model. The thick solid lines shows exact results obtained with the TI method, while the other lines show results obtained from Eqs. (3.14) and (3.15) with $\gamma = 30$, $p_{max} = 4$, and four different values of $n_{max}$ ($n_{max} = 30$, 40, 50, and 60).

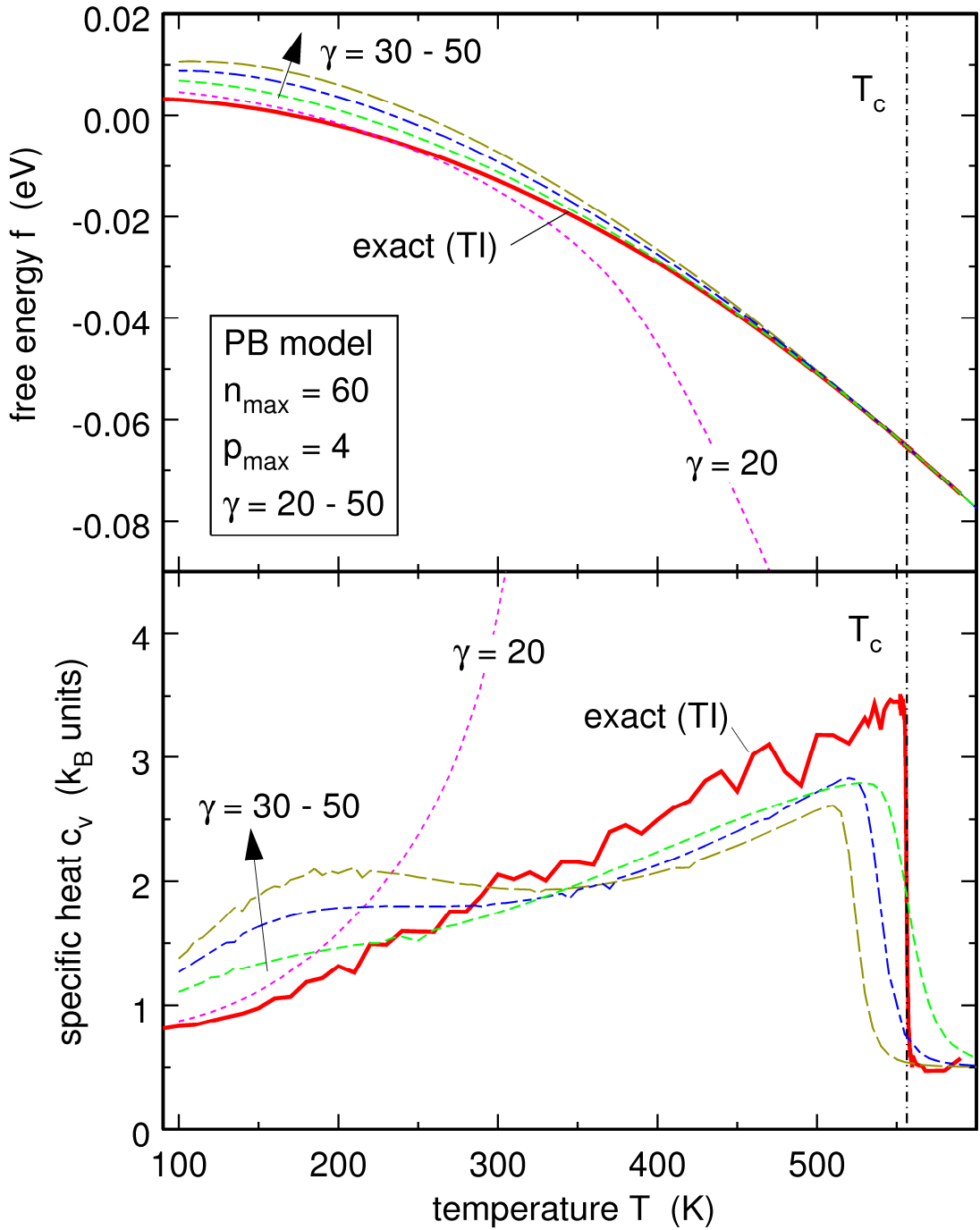

**Fig. 3** : Same as Fig. 2, but for $n_{max} = 60$, $p_{max} = 4$, and four different values of $\gamma$ ($\gamma$=20, 30, 40, and 50).

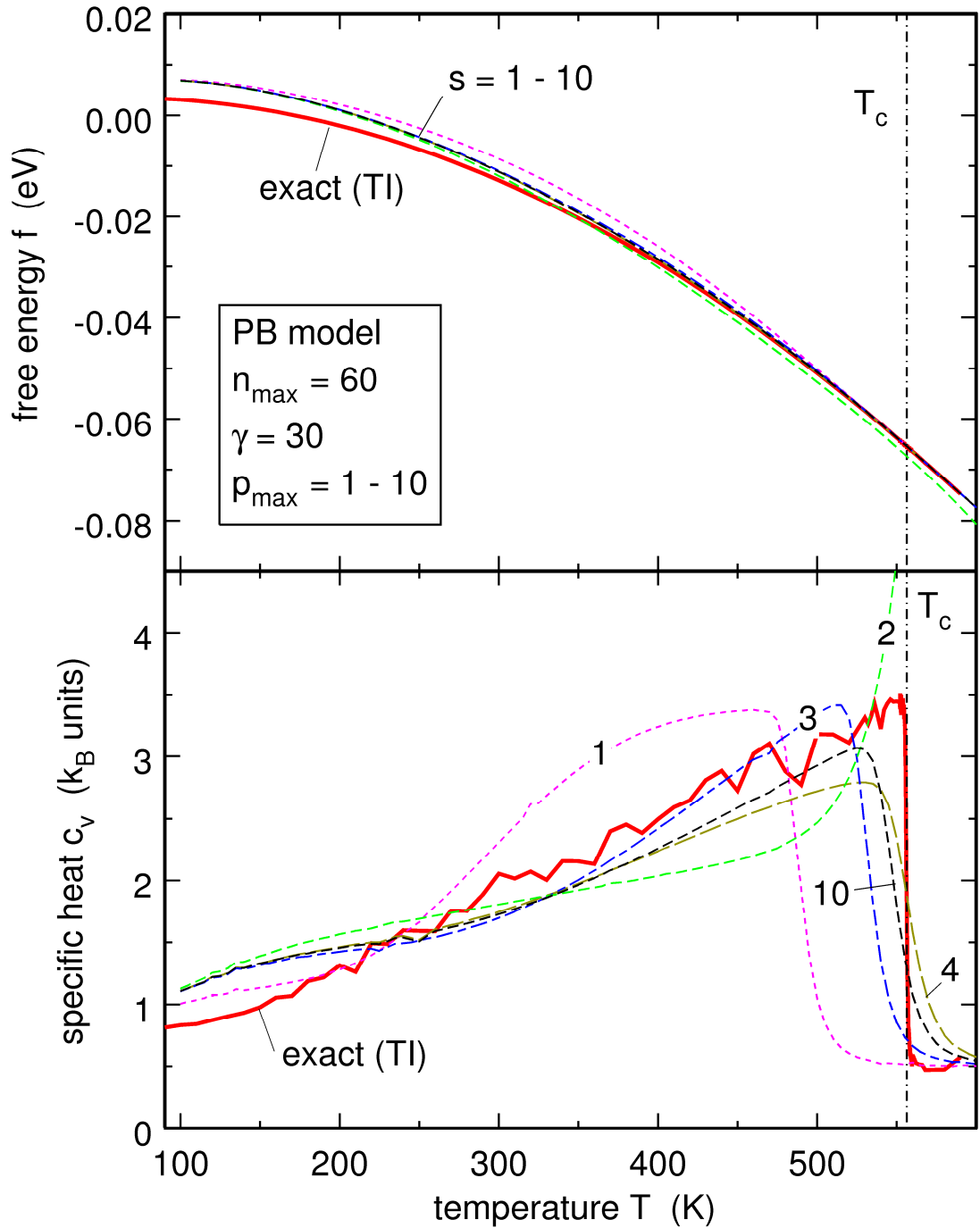

**Fig. 4** : Same as Fig. 2 but for $n_{max} = 60$, $\gamma = 30$, and six different values of $p_{max}$ ($p_{max}$ =1, 2, 3, 4, 10 and 20). The curves with $p_{max}$ =10 and $p_{max}$ =20 superpose.

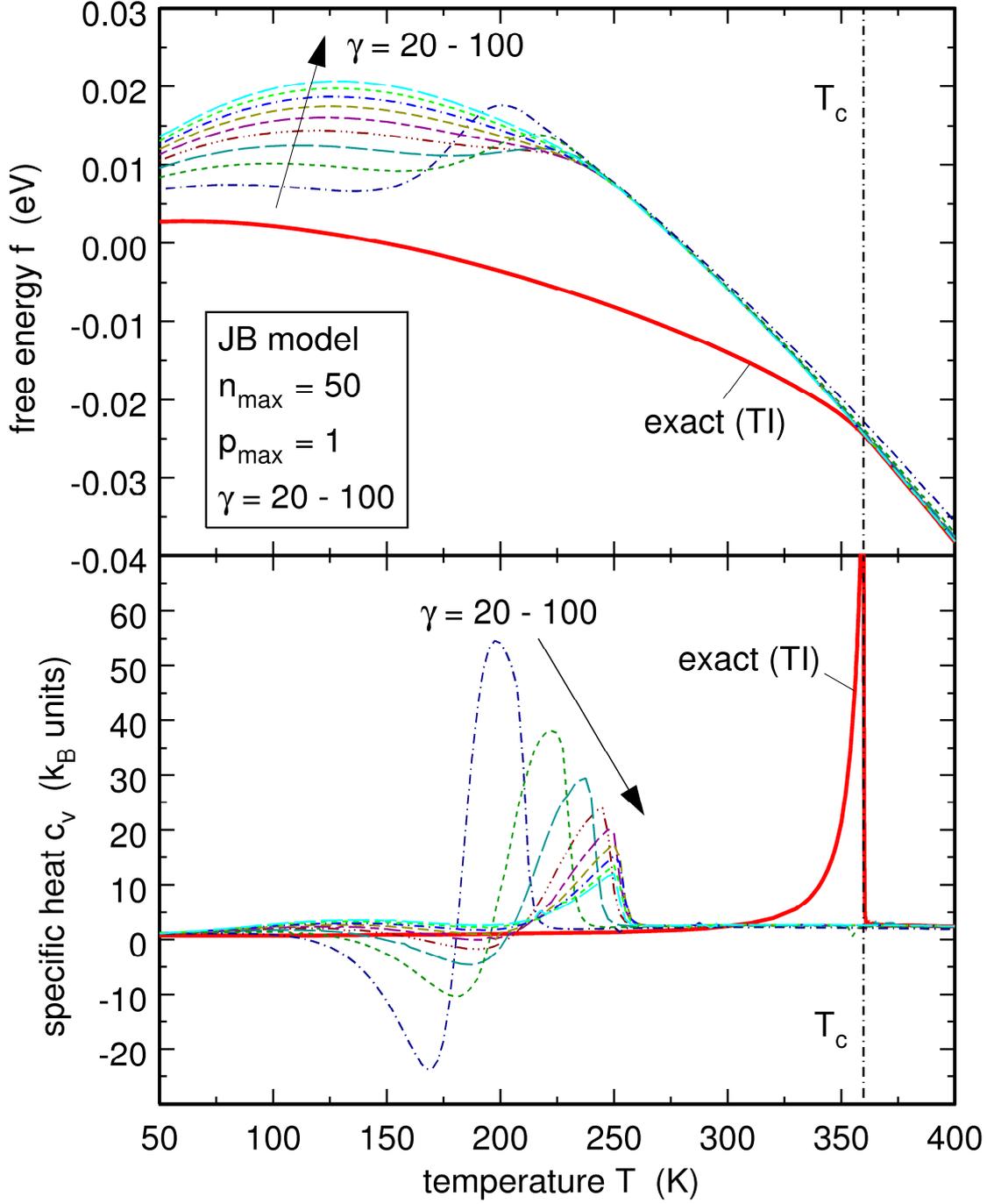

**Fig. 5** : Temperature evolution of $f$, expressed in eV (top plot), and $c_V$, expressed in units of $k_B$ (bottom plot), for the JB model. The thick solid lines shows exact results obtained with the TI method, while the other lines show results obtained from Eqs. (3.14) and (3.15) with $n_{max} = 50$, $p_{max} = 1$, and nine values of $\gamma$ regularly spaced between $\gamma=20$ and $\gamma=100$.

FIGURE 6

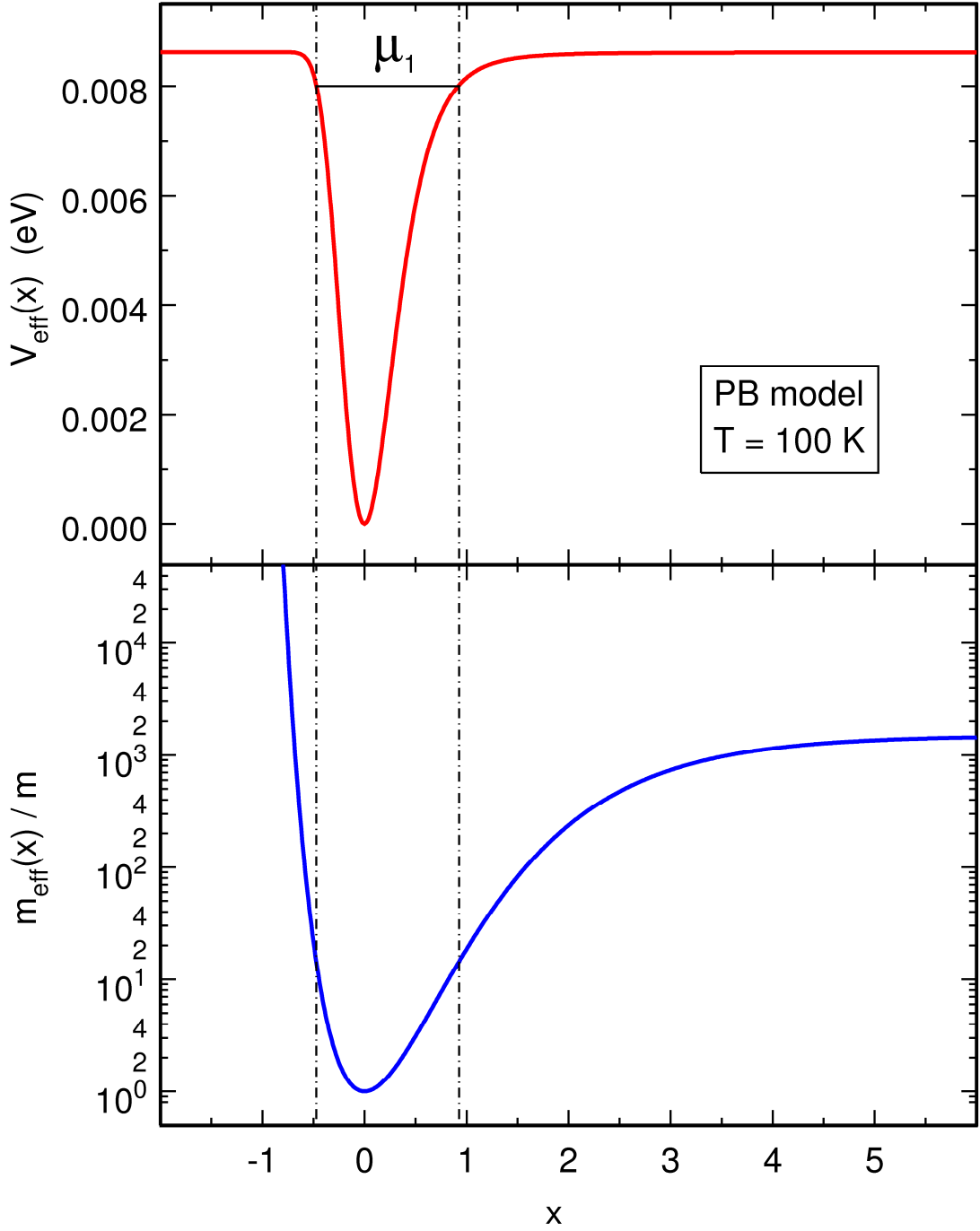

**Fig. 6** : Plots of $V_{eff}(x)$, expressed in eV (top plot), and $m_{eff}(x)/m$ (bottom plot), for the PB model at $T$=100 K. $V_{eff}(x)$ and $m_{eff}(x)$ are defined in Eq. (5.2). The horizontal segment in the top plot shows the energy $\mu_1$ of the lowest eigenstate of the generalized PSE in Eq. (5.1).

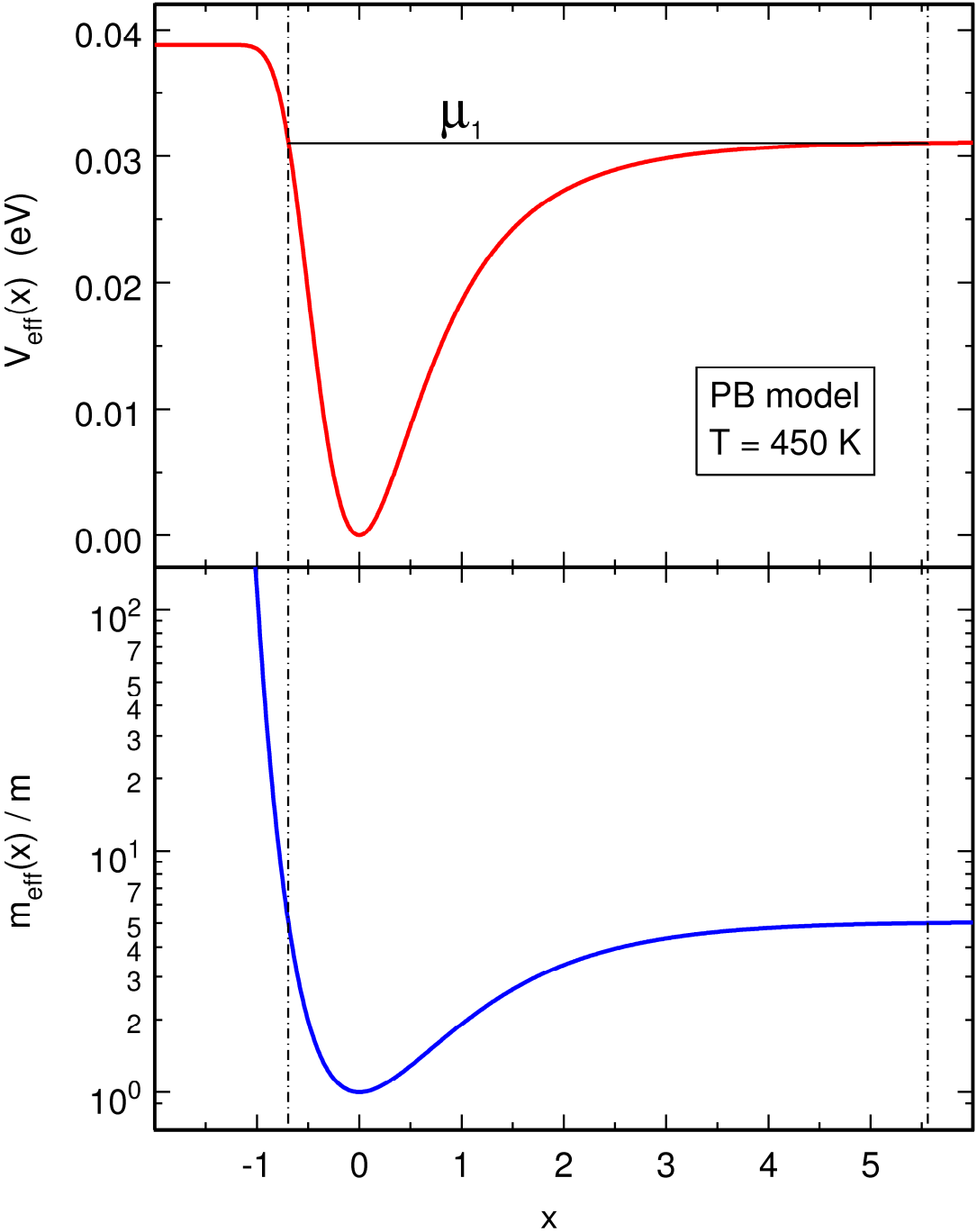

**Fig. 7** : Same as Fig. 6, but for *T*=450 K.